\documentclass{JHEP3}

            \input{epsf}
            \usepackage{epsfig}

         \usepackage{amssymb}
         \usepackage{latexsym}
         \usepackage{amsmath}

         \newcommand\fverb{\setbox\pippobox=\hbox\bgroup\verb}
         \newcommand\fverbdo{\egroup\medskip\noindent%
         \fbox{\unhbox\pippobox}\ }
         \newcommand\fverbit{\egroup\item[\fbox{\unhbox\pippobox}]}
         \newbox\pippobox

         \newcommand{\co}{{\cal O}{}}
         \newcommand{\cf}{{\cal F}{}}

         \newcommand{\ca}{{\cal A}{}}
         \newcommand{\cn}{{\cal N}{}}
         \newcommand{\cs}{{\cal S}{}}
         
         \newcommand{\ch}{{\cal H}{}}
         
         \newcommand{\cp}{{\cal P}{}}
         
         \newcommand{\ls}{{\frak{sl}(2,\mathbb{R})}{}}
         \newcommand{\SL}{\mbox{SL}(2,\mathbb{R})}
         \newcommand{\EBTZ}{\widetilde{\rm BTZ}}
         \newcommand{\AN}{\ca\,\cn}
         \newcommand{\dirac}{\slash {\hspace{-2.5mm} D}}
         \newcommand{\Ddirac}{\slash {\hspace{-2.5mm} \mathbb D}}
         \newcommand{\ads}{{\mbox{AdS}_3}{}}
         \newcommand{\h}{\,{\bf H}}
         \renewcommand{\t}{\,{\bf T}}
         \newcommand{\e}{\,{\bf E}}
         \newcommand{\f}{\,{\bf F}}
         \newcommand{\s}{\,{\bf S}}

         \newcommand{\x}{\,{\bf X}}
         \newcommand{\y}{\,{\bf Y}}
         \newcommand{\z}{\,{\bf Z}}

         \newcommand{\bx}{\,{\bf x}}
         \newcommand{\bz}{\,{\bf z}}
         \newcommand{\be}{\,{\bf e}}
         \newcommand{\ZZ}{\mathbb Z}
         \newcommand{\CC}{\mathbb C}
         \newcommand{\RR}{\mathbb R}
         \newcommand{\rr}{{\mbox{{\rm I}\kern-.2em\hbox{\rm R}}}}
         \newcommand{\UU}{\mathbb H}
         
         \newcommand{\Id}{\mbox{\rm \bf 1\hspace{-1.2 mm}I}}
         \def\st#1#2{\underset{#2}{\overset{#1}{\star}}}

         \newcommand{\tr}{\mbox{\rm Tr\,}}
         \newcommand{\fin}{\end{document}}
         \def\a#1#2{\alpha_{\bf #1\,\bf #2}}
         \def\un#1#2{\nu_{\bf #1\,\bf #2}}

         \def\lambdabar{{\mathchar'26\mkern-12mu\lambda}}
         \def\rpartial{\mathrel{\partial\kern-.75em\raise1.75ex\hbox{$\rightarrow$}}}
         \def\lpartial{\mathrel{\partial\kern-.75em\raise1.75ex\hbox{$\leftarrow$}}}
         \def\b#1{{\bf #1}}

         \newcommand{\R}{\mathbb R}

         \newcommand{\cb}{{\cal B}{}}
         \newcommand{\ce}{{\cal E}{}}

         \title{Star products on extended massive non-rotating BTZ black holes}

         \author{P.~Bieliavsky \\{\it Service de G\'eom\'etrie diff\'erentielle}\\ {\it
         Universit\'e Libre de Bruxelles, Campus Plaine, C.P. 218}\\ {\it
         Boulevard du Triomphe, B-1050 Bruxelles, Belgium}\\ E-mail:
         \email{pbiel@ulb.ac.be}}
         \author{S.~Detournay \thanks{"Chercheur FRIA", Belgium}~ and Ph.~Spindel
         \\{\it M\'ecanique et Gravitation}\\ {\it Universit\'e de Mons-Hainaut, 20
         Place du Parc}\\ {\it 7000 Mons, Belgium}\\E-mail:
         \email{stephane.detournay@umh.ac.be},\email{spindel@umh.ac.be}}
         \author{M.~Rooman \thanks{FNRS Research Director}\\{\it Service de Physique
         th\'eorique}\\ {\it Universit\'e Libre de Bruxelles, Campus Plaine, C.P.225}\\
         {\it Boulevard du Triomphe, B-1050 Bruxelles, Belgium}\\E-mail:
         \email{mrooman@ulb.ac.be}}

         \received{} 
         \revised{}
         \accepted{} 

         \abstract{$AdS_3$ space-time admits a foliation by two-dimensional
         twisted conjugacy classes, stable under the identification
         subgroup yielding the non-rotating massive BTZ black hole. Each
         leaf constitutes a classical solution of the space-time
         Dirac-Born-Infeld action, describing an open D-string in $AdS_3$
         or a D-string winding around the
         black hole. We first describe two nonequivalent maximal extensions
         of the non-rotating massive BTZ space-time and observe that in one
         of them, each D-string worldsheet admits an action of a
         two-parameter subgroup ($\ca \cn$) of $\SL$. We then construct non-formal,
         $\ca \cn$-invariant, star products that deform the
         classical algebra of functions on the D-string worldsheets and on their
         embedding space-times. We end by giving the
         first elements towards the definition of a Connes spectral triple on
         non-commutative $AdS$ space-times.}


         \dedicated{Dedicated to\ldots\\Marguerite.}

\begin{document}

         \section{Introduction} Since its
         discovery \cite{DT}, the BTZ black hole solution of 2+1
         dimensional gravity has proven to be a useful tool for exploring
         fundamental issues in black hole physics, both at the classical and
         quantum levels. Perhaps one of the most remarkable properties of the
         BTZ solution comes from its connection with string theory. On the one hand,
         the BTZ
         black hole can be represented as a quotient of the group manifold
         $\widetilde{\SL} \simeq AdS_3$ by a discrete subgroup of
         its isometry group
         $\widetilde{\SL} \times \widetilde{\SL}$ \cite{BTZ,BHTZ}. On the
         other hand, a $\widetilde{\SL}$ WZW model is an exact string
         theory background, describing the propagation of strings on the
         group manifold. Hence, quotienting out the appropriate
         discrete subgroup in this model leads to a theory that corresponds to an
         {\itshape exact} string theoretical representation of the BTZ
         black hole \cite{Str,Kal}.

         The BTZ space-time reveals a rich geometric structure.
         In the generic case ($0\leq
         J < M$), $AdS_3$ space-time admits a global foliation by two
         dimensional leaves stable under the action of the identification
         subgroups leading to the black hole \cite{BRS,BDHRS}.
         Furthermore, the universal
         covering space of a generic BTZ space-time realized as some open domain
         in $\ads$ is, in a canonical
         way, the total space of a principal fibration over $\R$ with as
         structure group a minimal parabolic subgroup $\ca \cn$ of
         $\widetilde{SL(2,\R)}$. The action of the structure group is
         isometric with respect to the $\ads$ metric on the total space. Moreover,
         each of the fibers is canonically endowed with a Poisson
         structure, constituting the primary input for the construction of
         star products, i.e. deformations of the pointwise multiplication
         of functions defined on the leaves.

         For spinless BTZ black holes, the stable leaves are twisted
         conjugacy classes in $\widetilde{SL(2,\R)}$ \cite{BRS}. Such
         classes are known to be WZW branes in
         $\ads$ \cite{FluxStab,Stanciu,BachPetr}, more precisely, they are
         extremal for
         the Dirac-Born-Infeld (DBI) brane action associated to a specific
         2-form $B$ on $\ads$
         (referred hereafter as the `$B$-field'). Consequently, they may be
         interpreted as {\itshape closed D1-branes (D-strings)} winding
         around the black hole.

         For this reason, we restrict ourselves in the present work to the
         non-rotating massive BTZ black hole. We define
         deformations of the algebra of functions on BTZ spaces,
         supported on (or tangential to) the leaves, and
         require these deformations to be non-formal (`strict' in the sense
         of Rieffel, see below) and compatible with the action of
         the structure group $\ca \cn$. Our motivation is threefold.
         First, a deformation of the brane in the direction of the
         $B$-field is generally understood as the (non-commutative)
         geometrical framework for studying interactions of strings with
         endpoints attached to the brane \cite{BigSuss,SchomD,SeibWitt}.
         Though curved non-commutative situations have been extensively
         studied in the context of strict
         deformation theory {\it i.e.} in a purely operator algebraic
         framework (for a review and
         references, see \cite{BranesCurved}), non-commutative spaces
         emerging from string theory have
         up to now mainly been studied in the case of constant $B$-fields in flat
         (Minkowski) backgrounds.

         The second motivation relies on the work of Connes and Lott \cite{ConLot},
         who used Rieffel's strict deformation method for actions of
         tori to define spectral triples for non-commutative spherical
         manifolds. The main point of their construction is that the data
         of an {\sl isometric} action of a torus on a spin manifold yields
         not only a strict deformation of its function algebra but a
         compatible deformation of the Dirac operator as well. Therefore,
         obtaining a strict deformation formula for actions of $\ca \cn$,
         in the context of BTZ spaces, would yield an example of
         spectral triple for non-commutative non-compact Lorentzian
         manifolds with constant curvature, with the additional feature that
         the deformation would be supported on the $\ca \cn$-orbits. The
         difficulty here is of course that these orbits cannot be obtained
         as orbits of an isometric action of $\R^d$.

         Finally, as the maximal fibration preserving isometry group (the
         universal cover of) of a BTZ space is the minimal
         parabolic group $\ca \cn$, it is natural to ask for a deformation
         which is invariant under the action of $\ca \cn$. But there is also a
         deeper reason for requiring this invariance. On the basis of considerations on
         black hole entropy or just by geometrical interest, one could be tempted
         to define higher genus locally $\ads$ black holes
         (every BTZ black hole is topologically $S^1\times\R^2$). This
         of course implies implementing the action of a Fuschian group
         in the BTZ picture. At the classical level, this possibility has
         already been investigated \cite{Brill}. Our
         point here is that it may also be investigated at the
         deformed level. Indeed, the symmetry group $\ca \cn$ is certainly too
         small to contain large Fuschian groups, but if the
         deformation is already $\ca \cn$-invariant, the
         (classical) action of $\ca \cn$ could perhaps be extended to a
         (deformed) action
         of the entire $\ads$ by automorphisms of the deformed algebra. If
         this is the case (and it is), an action of every
         Fuschian group is obtained at the deformed functional level.
         These issues will however not be tackled in the present work.

         This paper is organized as follows. In section \ref{Geom}, we
         recall some geometrical properties of the
         non-rotating massive BTZ black hole, and describe two
         nonequivalent maximal extensions of the BTZ space-time. We
         focus on one of the extensions, where each leaf of the foliation
         admits an action of $\ca \cn$. In section \ref{Branes}, we study the
         space-time properties of the winding D-strings in the extended BTZ space
         using the DBI action. Section 4 is devoted
         to star products. We first discuss general properties of invariant
         star products on group manifolds and how they induce star products
         on manifolds admitting an action of this group. We then construct
         a family of star products on the group $\ca \cn$ and on the winding
         D1-branes' worldvolumes in $AdS_3$ and BTZ space-times.
         In section \ref{Dirac}, deformed Dirac operators are defined
         in view of obtaining Connes' spectral triples. Section 6 contains
         conclusions and perspectives.

         \section{Geometry of the extended BTZ black holes}\label{Geom} We
         have previously shown
         \cite{BRS} that BTZ black holes are canonically endowed with a
         regular Poisson structure, admitting a characteristic foliation
         constituted by the orbits of an external bi-action. We shall
         briefly summarize this construction. The Lorentzian space $AdS_3$
         is defined as the universal covering of the group:
         \begin{equation}
         \label{matbz} \SL=\{\bz = \left(
         \begin{array}{cc} u+x & y+t\\ y-t & u-x
         \end{array}
         \right)| x,y,u,t\in\mathbb{R},\ \det \bz =1 \} \qquad .
         \end{equation} Its Lie algebra
         \begin{equation}
         \label{matsl} \ls=\{\left(\begin{array}{cc} z^H & z^E \\ z^F &
         -z^H
         \end{array}\right):= z^H\h+z^E\e+z^F\f\} \qquad ,
         \end{equation} is expressed in terms of the generators $\{\h , \e, \f
         \}$ satisfying the commutation relations:
         \begin{equation}
         \label{comrel} [\h ,\e]=2\e \quad , \quad [\h,\f]=-2\f \quad ,
         \quad [\e,\f]=\h \quad .
         \end{equation} By identifying $\ls $ with the tangent space of $\SL$ at the
         identity element $\be$, the Killing metric at this point, denoted
         by $\beta_{\be}$, is given by:
         \begin{equation}
         \beta_{\be}(\x,\y):= {\frac 1 2} \tr(\x \y)\qquad ;\qquad \x \, ,
         \y \in \ls\ .
         \end{equation} The Lorentzian character of $AdS_3$ means that there is an
         isomorphism between $\ls$ and $M^{2,\, 1}$, the Minkowski space in
         2+1 dimensions. The generator $\h$ is space-like whereas $\e$ and
         $\f$ are light-like, since:
         \begin{equation}
         \beta_{\be} (\h\, ,\h)=1 \qquad ,\qquad \beta_{\be} (\e\, ,\f)=1/2
         \qquad ,\qquad \beta_{\be} (\e \, ,\e)=0=\beta_{\be} (\f\, ,\f)
         \qquad .
         \end{equation} Let us also introduce the unit generator $\t=\e - \f$, and the
         $\SL$ subgroups:
         \begin{equation} {\cal A}=\exp(\RR\,\h)\ ,\quad {\cal{N}}=\exp(\RR\,\e)\
         ,\quad {\cal{K}}=\exp({\mathbb I}\,\t)\ ,\ \qquad {\mathbb I}=[ 0
         \, , 2\, \pi ]\qquad ,
         \end{equation} which are the building blocks of the Iwazawa decomposition of
         $\SL={\cal K\,A\,N}$.

         The automorphism group of $\ls$ is isomorphic to the
         three-dimensional Lorentz group $SO(2,\, 1)\equiv
         L^\uparrow_+(2,\, 1) \cup L^\downarrow_-(2,\, 1) $.
         Transformations belonging to $L^\uparrow_+(2,\, 1)$ correspond to
         internal isomorphisms; those belonging to $L^\downarrow_-(2,\, 1)$
         need the introduction of an external automorphism that we choose
         as:
         \begin{equation}
         \sigma(\h)=\h\qquad,\qquad\sigma(\e)=-\e\qquad,\qquad\sigma(\f)=-\f\qquad,
         \end{equation} or equivalently, using the matrix representation (\ref{matsl}):
         \begin{equation}
         \sigma(\z)=\h\, \z\, \h\qquad,\qquad \z\in\ls \qquad .
         \end{equation}

         A massive non-rotating BTZ black hole is obtained as the
         equivalence classes of the quotient of $\SL$\strut\footnote{We may
         restrict ourselves to $\SL$ as we focus on non-rotating black
         holes; for rotating black holes, $AdS_3 =\widetilde{\SL}$ has to
         be considered; see ref. \cite{BDHRS} for details.} under the
         bi-action of the subgroup $\ch $, defined by
         \begin{equation}\label{biaction}
         \ch \times \SL\rightarrow \SL:({\bf h},\ {\bf z})\mapsto {\bf
         h}\,{\bf z}\, \sigma({\bf h}^{-1}):=\Sigma^\sigma_{\bf h}({\bf z})
         \end{equation} where the subgroup $\ch$ is the restriction of $\ca$ to integer
         values of its parameter:
         \begin{equation}\label{BHTZsg}
         \ch = \exp(\pi \sqrt{M} \, \mathbb Z \h)\qquad ,
         \end{equation} where $M$ is the mass of the black hole.

         The $\SL$ group can also be viewed as the quadric $Q\equiv
         t^2+u^2-x^2-y^2=1$ in the four-dimensional flat ultra-hyperbolic
         space $M^{2,\, 2}$, with metric $ds^2=-dt^2-du^2+dx^2+dy^2$, while
         the orbit of a point $z$ under the $\SL$ bi-action:
         \begin{equation}\label{orbit}
         \co_z =\{\b g\,z\,\sigma(\b {g^{-1}})\vert\, \b g\in \SL\}
         \end{equation} is given by the intersection of $Q$ with the planes of constant
         $x$ coordinate. Accordingly, these orbits are two-dimensional
         hyperboloids, isomorphic to the unit hyperboloid $\UU=\{\x\in \ls
         \vert \beta_{\be}(\x,\,\x)=1 \}$ of $\ls$. The group $\SL$ acts on
         $\UU$ by the adjoint action. This allows to associate to each
         point $\x$ of $\UU$, expressed as $\x={\bf g}_0\h\,{\bf
         g}_0^{-1}$, the equivalence class of matrices:
         \begin{equation} [{\bf g}_0]=[\ \exp(\RR\,\x)\, {\bf g}_0\,
         \exp(\RR\,\h)\ ]=[{\bf g}_0\, \exp(\RR\,\h)\ ]\qquad .
         \end{equation} Now it is easy to check that we may represent $Q$, {\it i.e.}
         $\SL$, as the product $\UU\times \RR$ by considering the union of
         the orbits $\co_{\exp({{\rm I}\kern-.18em\rm R} \,\h)}$. Indeed,
         elementary calculations show that ${\bf
         g}_1\exp(\rho_1\,\h)\sigma({\bf g}_1^{-1})={\bf
         g}_2\exp(\rho_2\,\h)\sigma({\bf g}_2^{-1})$ implies that
         $\rho_1=\rho_2$ and $[{\bf g}_1]=[{\bf g}_2]$.

         The transformations defined by eqs (\ref{BHTZsg}),(\ref{biaction})
         constitute a discrete isometry subgroup of $\SL$, first introduced
         in ref. \cite{BHTZ}, and that hereafter we call the BHTZ subgroup.
         Let us emphasize that the orbits \ref{orbit} are stable under the
         action of the BHTZ subgroup. This remark, with the isomorphism
         $\SL\simeq \UU\times \RR$, are the key ingredients to obtain a
         global coordinate system on BTZ black holes. The generator of the
         transformation $\Sigma^\sigma_{\exp(\RR\,\h)}(z)$ partitions
         $AdS_3$ into different connected components according to its
         nature. On two connected regions, the generator is time-like. We
         denote these domains ${\bf I}$ and ${\bf I'}$. Moreover, there are
         four connected regions where the generator is space-like. They are
         denoted by ${\bf II_R}$, ${\bf II_L}$ and ${\bf III_R}$, ${\bf
         III_L}$ (see Fig.1). On the boundaries between these domains, the
         generator is light-like or zero. The causally safe domains ${\bf
         I}$ and ${\bf I'}$ correspond to $t(z)^2-y(z)^2
         >0$ \cite{BRS}. Expressing $z$ as $\Sigma^\sigma_{\b g}(\exp (\rho\,\h))$
         for suitably chosen $\b g$ and $\rho$, we may express this
         condition in terms of $\b g$ as:
         \begin{equation} 1>t(\b {g})^2-y(\b {g})^2>0\qquad .
         \end{equation} Let us notice that this last condition is
         $\rho$-independent, as expected. Writing $\b X=\b g\,\h\,\b
         g^{-1}=x^H \h + x^E \e + x^F \f $, this condition implies that:
         \begin{equation}
         \vert x^H\vert <1\qquad .
         \end{equation} In ref. \cite{BRS} we used the following parameterization :
         \begin{equation} \b X=\exp(\frac\theta 2\h)\exp(-\frac\tau 2\t)\,\h\,
         \exp(\frac\tau 2\t)\exp(- \frac\theta 2\h)
         \end{equation} leading to the expression of the metric on ${\bf I}$ (or ${\bf
         I'}$)
         \begin{equation}
         \label{metBRS} ds^2= L^2 \left ( d\rho^2+\cosh(\rho)^2
         (-d\tau^2+\sin^2\tau\,d\theta^2 ) \right ) \qquad ,
         \end{equation} where we introduced the length scale parameter $L$ related to
         the cosmological constant $\Lambda$ by $\Lambda= - L ^{-2}$. The
         boundary of a connected component of this domain is given by the
         surfaces $\tau=0,\ \tau=\pi$ or $\tau=\pi,\ \tau=2\pi $, which
         correspond to coordinate singularities due to the occurrence of
         closed light-like curves. Hereafter we restrict ourselves to the
         domain ${\bf I}$ where $\tau$ varies from $0$ to $\pi$.

         The singularities we encounter here are of the type of those
         described by Misner \cite{Mis}. On BTZ space, indeed, we also have
         two families of null geodesics that spiral as they approach the
         chronological horizons, located at $\tau=0$ and $\tau =\pi$, also
         called the BTZ
         black hole singularities \cite{BHTZ}. As
         explained in \cite{HE}, we may extend BTZ space across the
         singularities by selecting one of the families of null geodesics
         near each component of the singularities. According to the family
         of null geodesics that is untwisted, it is possible to extend a
         causally safe domain in four different ways; these extensions are
         two by two isomorphic. $AdS_3$ space-time can be viewed as a stack
         (i.e. a trivial fiber bundle) of leaves $\rho$=cst. On Fig.1, we
         have depicted the Penrose diagram of such a leaf and its
         intersection with the different regions ${\bf I}$, ${\bf I'}$,
         ${\bf II_R}$, ${\bf II_L}$, ${\bf III_R}$, ${\bf III_L}$. We have
         also represented the intersection of a leaf with a fundamental
         domain with respect to the action of the BHTZ subgroup. The two
         nonequivalent extensions of the BTZ black hole consist of the
         continuation of the central region ({\bf I}) by the two right
         regions $\mbox{\bf II}_{\bf R}$ and $\mbox{\bf III}_{\bf R}$ , (or
         equivalently the two left ones $\mbox{\bf II}_{\bf L}$ and
         $\mbox{\bf III}_{\bf L}$) or by extending the central region via a
         left and a right region. Despite the pathologies in its causal
         structure, the former type of extension is particularly
         interesting because it allows to define the action of an ${\cal
         A}\ {\cal N}$ group on each leaf of the foliation. Let us
         consider, to fix the idea, the domain obtained by the union
         \begin{equation}
         \b U={\bf I}\cup {\bf II}_R\cup{\bf III}_R\label{UR}
         \end{equation} We may parametrize $\b X$ as:
         \begin{equation}\label{Xext}
         \b X=\exp(\frac\phi 2\h)\exp( w\, \e)\,\s\, \exp(- w
         \,\e)\exp(-\frac\phi 2\h)
         \end{equation} where:
         \begin{equation}
         \s=\exp(-\frac\pi 4\t)\,\h\, \exp(\frac\pi 4\t)
         \end{equation} which leads to the maximally extended metric on the
         domain ${\bf U}$ :
         \begin{equation}
         \label{metmax} ds^2=L^2 \left ( d\rho^2+\cosh(\rho)^2
         (d\phi^2-(w\,d\phi+dw)^2) \right ) \qquad .
         \end{equation}
         Restricting $\phi$ to $[0,\, 2\pi\, \sqrt{M}]$ in (\ref{Xext}) and
         (\ref{metmax}), we obtain a maximally extended spinless BTZ black
         hole, denoted hereafter by $\EBTZ$. The chronological horizons
         (which are the BTZ "singularities" and correspond to Cauchy
         horizons in $AdS_3$) are located at $w=\pm 1$; the black hole
         horizons are given by
         \begin{equation}\label{horizons}
         \tanh \rho/2 = \pm \left(\frac{1-\sqrt{1-w^2}}{w} \right) \quad .
         \end{equation}
         On the intersection
         of the domains they cover, the coordinate systems (\ref{metBRS})
         and (\ref{metmax}) are related by
         \begin{equation} \label{changecoord}
         w = \cos \tau \quad \mbox{and} \quad e^{2 \phi} = \frac{e^{2
         \theta}}{\sin \tau} \, .
         \end{equation}

       \begin{figure}[hbt]
\begin{center}

\resizebox{!}{0.9\hsize}{\includegraphics{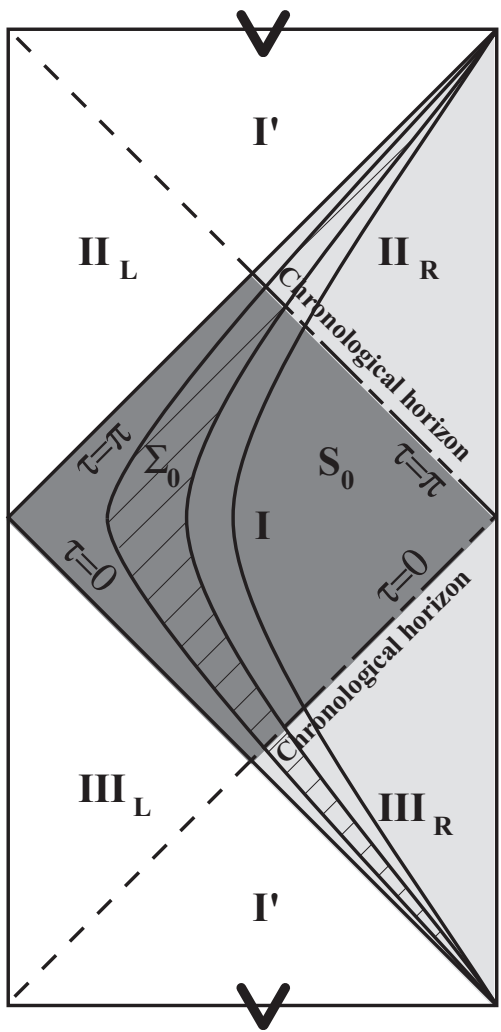}}
\caption[]{\label{EBTZ} This figure, with the top and bottom lines
identified, is a Penrose diagram of a $\rho=\rho_0$ section in
$\SL$. $AdS_3$ can be seen as a stack of such fibers (without this
identification). Accordingly, this diagram can also be seen as a
Penrose diagram of $\SL$, each point representing a line
parametrized by $\rho$, running from $-\infty$ to $+\infty$. The
region ${\bf I}$ (darkly shaded) provides after identification by
the BHTZ subgroup (eq. (\ref {BHTZsg})) the usual non-rotating BTZ
black hole space-time, bounded by the chronological horizons
$\tau=0$ and $\tau=\pi$. The maximally extended region ${\b S_0}$
(shaded) admits an action of $\ca \cn$, and represents the
intersection of the domain $\b U$ defined in (\ref{UR}) with a
$\rho=\rho_0$ section. This maximal extension goes beyond the
chronological horizons, where the identifications yielding the
black hole become light-like, and which usually referred as the
BTZ black hole {\itshape chronological singularity} (the BTZ black
hole singularity is not a curvature singularity, but merely a
singularity in the causal structure). The dashed region $\Sigma_0$
represents a fundamental domain of the action of the BHTZ
identification subgroup on ${\b S_0}$, i.e. a $\rho=\rho_0$
section of the extended $\EBTZ$ black hole.}

\end{center}
\end{figure}

         \section{Symmetric D-branes in $\EBTZ$ space-time}\label{Branes}
         Strings moving in group
         manifolds are described by the WZW model \cite{Gep-Witt}. The
         exact conformal invariance of these models is based on the current
         algebra symmetry, generated by the Lie algebra valued chiral
         currents $J(z)$ and $\bar{J}(\bar{z})$. A well-understood class of
         D-brane configurations in WZW models is obtained as solutions of
         the familiar gluing conditions on the chiral currents
         \begin{equation} J(z) = R \bar{J}(\bar{z})
         \end{equation} at the boundary of the string worldsheet, where $R$ is a metric
         preserving Lie algebra automorphism \cite{Stanciu,AlexSch}. These
         gluing conditions describe symmetric D-branes, that is,
         configurations which preserve conformal invariance and the
         infinite-dimensional symmetry of the current algebra of the bulk
         theory . The geometry of the associated branes is encoded in these
         gluing conditions. Their worldvolumes are shown to lie on
         (twisted) conjugacy classes in the group manifold and constitute
         classical solutions of the space-time DBI
         action \cite{FluxStab}.

         Symmetric D-branes of the $Sl(2,\RR)$ WZW model are of three types
         : two-dimensional hyperbolic planes ($H_2$), de Sitter branes
         ($dS_2$) and anti-de Sitter branes ($AdS_2$). In \cite{BachPetr},
         it was shown that the $AdS_2$ worldvolumes, corresponding to
         twisted conjugacy classes, are the only physically relevant
         classical configurations, solutions of the DBI equations. Indeed,
         the worldvolume electric field on the $dS_2$ branes is
         supercritical, while $H_2$ branes have euclidean signature and
         must therefore be interpreted as instantons.

         We now make a link with the geometry of the non-rotating BTZ
         black-hole. We saw in the previous section that the spinless BTZ
         black hole admits a foliation by leaves, the $\rho$ = constant
         surfaces, which are stable under the action of the BHTZ subgroup
         and constitutes twisted conjugacy classes in $Sl(2,\RR)$ ($AdS_2$
         spaces). From our previous discussion, each of these leaves
         constitutes a D1-brane that is, a solution of the equations of
         motion derived from the DBI action, which constitutes the
         effective action for Dp-branes\footnote{We have not considered
         Wess-Zumino terms; they vanish when all Ramond-Ramond background
         fields are set to zero} :
         \begin{equation}\label{BI} S_{BI} = T_p \int d^{p+1}x \sqrt{- det(\hat{g} +
         \hat{B} + 2 \pi \alpha^{'} F)} = T_p \int d^{p+1}x \: L_{BI}
         \quad,
         \end{equation} where $\hat{g}$ and $\hat{B}$ are the pull-backs of the WZW
         backgrounds and $F$ is the worldvolume electric field. From now
         on, we set $2 \pi \alpha' =1$. With $K_{ij} = g_{ij} + B_{ij} +
         F_{ij}$ and $K = \mbox{det}(K_{ij})$, the equations of motion (for
         Abelian $F$-field) derived from the DBI action are :

         \begin{eqnarray}
         \partial_k(\sqrt{- K} K^{(kj)} X^{\mu}_{,j})g_{\mu \lambda} +
         \sqrt{-K} K^{(kj)} \Gamma_{\mu \nu,\lambda}
         X^{\mu}_{,j}X^{\nu}_{,k} + \sqrt{-K} K^{[kj]} H_{\mu \nu \lambda}
         X^{\mu}_{,j}X^{\nu}_{,k} = 0
         \nonumber\\
         \partial_i(\sqrt{- K} K^{[ij]}) = 0 \label{eqA} \quad,
         \end{eqnarray} with $K^{(kj)}= \frac{1}{2} (K^{kj} + K^{jk})$ and
         $K^{[kj]}=\frac{1}{2}(K^{kj} - K^{jk})$, $H_{\mu \nu \lambda}
         =\frac{1}{2}(B_{\mu \nu, \lambda}+B_{\nu \lambda,\mu}-B_{\mu
         \lambda,\nu})$ and $\Gamma_{\mu \nu,\lambda} = \frac{1}{2}(g_{\mu
         \lambda,\nu} + g_{\nu \lambda,\mu}-g_{\mu \nu \lambda})$.
         Furthermore, in presence of space-time Killing vectors $\Xi$, such
         that ${\cal L}_{\Xi} B = d \widetilde{\alpha}$ (a condition
         trivially satisfied here), the following current is conserved
         on-shell:
         \begin{equation}\label{intprem} J^i = \Xi^{\mu} \frac{\partial L}{\partial
         X^{\mu}_{,i}} - \alpha_j \frac{\partial L}{\partial A_{j,i}}
         \:,\quad \partial_i J^i \approx 0 \: ,
         \end{equation} with $\alpha_j = \widetilde{\alpha}_{\mu} X^{\mu}_{,j}$ .

         The WZW three-form $H$ is proportional to the volume form, the
         constant of proportionality being fixed by the condition of
         quantum conformal invariance. We obtain \footnote{The beta
         function equation reads as \cite{GSW} $R_{\mu \nu} = \frac{1}{4}
         H_{\mu}^{\lambda \omega} H_{\nu \lambda \omega}$. In our case, we
         have $R_{\mu \nu} = 2 \Lambda g_{\mu \nu}$, and fixing the
         space-time orientation, $k = 2$, see ref. \cite{BachPetr}.}
         \begin{equation} H_{\mu \nu \lambda} = 2 \: \sqrt{- g} \: \varepsilon_{\mu
         \nu \lambda} \quad.
         \end{equation}

         A globally defined metric on the extended BTZ space-time is given
         by (\ref{metmax}). We fix the gauge by choosing in these
         coordinates the B-field:
         \begin{equation} B = (\rho + \sinh \rho \cosh \rho) dw \wedge d\phi \quad .
         \end{equation} Solutions of the DBI equations(\ref{BI}) read as:
         \begin{eqnarray}\label{Solext}
         \rho (x_0,x_1) &=& \rho_0, \quad w (x_0,x_1) = x_0, \quad \phi
         (x_0,x_1) = x_1 \nonumber\\ F_{01}(x_0,x_1) &=& -\rho_0 \quad ,
         \end{eqnarray}where $x_0$ and $x_1$ are the worldvolume coordinates on the
         brane.

         These solutions correspond to projections of twisted $AdS_3$
         conjugacy classes that wrap around BTZ space, and are thus
         compatible with the restriction of the $\phi$-variable to its
         range $[0,2\pi\,\sqrt{M}]$. They may be interpreted as closed DBI
         1-branes in BTZ space. In general, the $AdS_3$ branes obtained
         from (\ref{Solext}) by action of isometries do not project into
         closed branes in BTZ, but into infinite branes that extend from
         $\rho=-\infty$ to $\rho=+\infty$. Only the $AdS_3$ branes obtained
         from isometries compatible with the identifications generated by
         the BHTZ subgroup lead to closed DBI branes in $\EBTZ$ space.
         These isometries correspond to the left and right action of the
         subgroup $\exp(\RR \h)$ on $\SL$. In terms of the coordinates
         (\ref{metmax}), their Killing vectors read as
         \begin{equation} \label{KillVec}
         \Xi_- = \partial_{\phi} \quad \mbox{and} \quad \Xi_+ = -w \,
         \partial_{\rho} - w \tanh \rho \, \partial_{\phi} + (w^2 -1)
         \tanh \rho \, \partial_{w} \,.
         \end{equation}
         The action of the corresponding isometries on the brane
         (\ref{Solext}) yields
         \begin{eqnarray} \label{Solextiso}
         \sinh \rho(x_0) &=& + x_0 \cosh \rho_0
         \sinh f_{\pm} +\sinh \rho_0 \cosh f_{\pm} \nonumber \\
         w^2 (x_0) &=& 1 + \frac{\cosh^2 \rho_0}{\cosh^2 \rho (x_0)} (x_0^2
         -1) \nonumber \\
         e^{2 \phi(x_0,x_1)} &=& \frac{\cosh^2 \rho (x_0)}{\cosh^2 \rho_0}
         \, e^{2 (g_{\pm} + x_1)} \, ,
         \end{eqnarray} where $f_{\pm}$ and $g_{\pm}$ are constants related to the
         isometries used to perform the transformations, $f_+ = g_- = 2\pi
         \, \sqrt{M}$ and $f_- = g_+ =0$. On the causally-safe region,
         these solutions are more easily expressed, using the coordinates
         of (\ref{metBRS}), as
         \begin{eqnarray} \label{Solregiso}
         \sinh \rho(x_0) &=& \cos x_0 \cosh \rho_0 \sinh f_{\pm} +
         \sinh \rho_0 \cosh f_{\pm} \nonumber \\
         \sin \tau(x_0) &=& \frac{\cosh \rho_0 \sin x_0}{\cosh \rho
         (x_0)} \nonumber\\
         \theta(x_1) &=& x_1 + g_{\pm} \quad ,
         \end{eqnarray} The corresponding expressions for the worldvolume electric
         field $F_{01}$ can then be deduced from (\ref{eqA}). The two
         constants of motion associated to the solutions (\ref{Solext}) and
         (\ref{Solextiso}) corresponding to the conserved currents $J_+$
         and $J_-$ [see (\ref{intprem}) and (\ref{KillVec})] are
         \begin{equation}
         J^0_- = 0 \quad \mbox{and} \quad J^0_+ = -\sinh \rho_0 \, .
         \end{equation}
         The isometries generated by $\Xi_-$ preserve globally the
         $\rho$=cst D1-brane worldvolume. The action simply consists in a
         rotation of the brane on itself. Those generated by $\Xi_+$
         generate boosts of the brane.

         \section{Star products}

         In this section we construct star products on the branes just obtained.

         \subsection{Induced star products}\label{IstP}
         Here we follow the method described in \cite{Pierre2,Pierre3}.
         Let us remind that a star product is
         a one-parameter (denoted $\lambdabar$) deformation of the algebra
         of functions on a (symplectic) manifold with pointwise
         multiplication, defining an associative composition law on a
         functional space that admits left and right unit elements, and
         such that to first order the deformed commutator provides the
         Poisson bracket.

         Consider a Lie group $G$. If $u$ is a complex valued function on
         $G$, {\it i.e.} $u \in {\rm Fun}[G,\CC]$, we denote by
         $L^\star_{\b g} [u]$ (resp. $R^\star_{\b g} $) the composition of
         this function with the left (resp. right) translations on the
         group:
         \begin{equation}
         \forall \ {\b g},\,{\b h}\in G,\quad L^\star_{\b h} [u](\b
         {g})=u({\bf h\,g})\quad,\quad R^\star_{\b h} [u]({\b g})=u({\bf
         g\,h})\qquad .
         \end{equation} If the group $G$ also acts on a space $X$, let say by a left
         action $\tau$:
         \begin{equation} G\times X\rightarrow\kern-1em^\tau\kern0.5em X\ : \ (\b
         g,\,\b x)\mapsto \tau_{{\b g}}(\b {x})\qquad \mbox{with}
         \qquad\tau_{{\bf g\,h}}=\tau_{\b g}\circ \tau_{\b h}\qquad ,
         \end{equation} we may use the induced action of $\tau$ on ${\rm Fun}[X,\CC]$
         to define an induced star product, $\ \st {}X$, on $X$, as
         follows. We denote by $\alpha_{\b g} $ the induced action of
         $\tau$, defined as
         \begin{equation}
         \alpha_{\b g} [u](\b {x})=u(\tau_{{\b
         g}^{-1}}(\b {x}))\quad , \forall \ \b{x}\in X, \quad \alpha_{{\bf
         g\,h}}=\alpha_{\b g}\circ \alpha_{\b h}\qquad .
         \end{equation} Accordingly, for fixed ${\b x}\in X$, we may define a map
         $\tilde\alpha^{\b x}$ from ${\rm Fun}[X,\,\CC]$ into ${\rm
         Fun}[G,\,\CC]$:
         \begin{eqnarray} {\rm Fun}[X,\,\CC]\rightarrow\kern-1.2em^{\tilde\alpha^{\b
         x}}\kern0.5em{\rm Fun}[G,\,\CC]:
         \ u \mapsto \tilde\alpha^{\b x}[u]&&\\
         \forall\ \b g\in G\ :\ \tilde\alpha^{\b x}[u](\b {g})=u(\tau_{{\b
         g}^{-1}}(\b {x})) &&.
         \end{eqnarray} Let us also assume that on $G$ we have a left invariant
         star product, denoted by~$\st LG$, {\it i.e.} a star product
         satisfying the relation
         \begin{equation} L^\star_{\b g} [u \st LG v]=L^\star_{\b g} [u]\st LG
         L^\star_{\b g} [v] \qquad .
         \end{equation}
         From the left invariant star product on $G$, we induce a star
         product on $X$, denoted $\st {\ \ }X$, defined as :
         \begin{equation}\label{babel}
         \left(u \st {\ \ }X v\right)(\b {x}):=\left(\tilde\alpha^{\b
         x}[u]\st LG \tilde\alpha^{\b x}[v]\right)(\b e)\qquad,
         \end{equation}
         $\b e$ denoting the identity element of $G$. If, instead of
         choosing to evaluate the star product on $G$ at the identity, we
         use the point $\b g$, we obtain a composition law related to the
         first one by the action of $\alpha_{\b g} $:
         \begin{equation}
         \alpha_{\b g} \left(u\st{\ \ }X v\right)(\b
         {x})=\left(\tilde\alpha^{\b x}[u]\st LG \tilde\alpha^{\b
         x}[v]\right)(\b g)\qquad , \end{equation} but which is associative
         only for ${\b g} = {\b e}$. The $\st {\ \ }X$ of two functions
         exists only when the r.h.s of (\ref{babel}) exists (see
         discussions on functional space in the following sections).

         On the other hand, using the inversion map
         \begin{equation} G\times X\rightarrow\kern-1em^i\kern0.5em X\ : \ \b g\mapsto
         i(\b {g}):={{\b g}^{-1}}\qquad ,
         \end{equation} and its induced action on ${\rm Fun}[G,\, \CC]$:
         \begin{equation}
         \forall {\b g}\in G,\quad i^\star [u](\b {g}):=u({\b
         g^{-1}})\qquad ,
         \end{equation} we may construct a right invariant star product on $G$ starting
         from the left one:
         \begin{equation} u\st RG v:=i^\star \left[i^\star [u]\st LG i^\star
         [v]\right]\qquad .\label{LtoR}
         \end{equation} The right invariance of $\st RG$ results immediately from the
         left invariance of $\st LG$ and the relations on ${\rm
         Fun}[G,\,\CC]$:
         \begin{equation} R^\star _{\b g} \left[i^\star [u]\right]=i^\star
         \left[L^\star _{{\b g}^{-1}}[u]\right]\quad,\quad L^\star _{\b g}
         \left[i^\star [u]\right]=i^\star \left[R^\star _{{\b
         g}^{-1}}[u]\right]\qquad .
         \end{equation} Indeed we obtain:
         \begin{equation} R_{\b g} ^\star [u]\st RG
         R_{\b g} ^\star [v]=i^\star\left[i^\star\left[R_{\b g}
         ^\star[u]\right]\st LG i^\star\left[ R_{\b g}
         ^\star[v]\right]\right]=R^\star_{\b g} \left[u\st RG
         v\right]\qquad .
         \end{equation}

         \subsection{ $\AN$-invariant star products} In section
         (\ref{Branes}), we have shown that
         the DBI branes $\rho =\rho_0$, in $AdS_3$ as well as in a
         maximally extended BTZ black-hole, are orbits of the action of an
         $\AN$ group. In this subsection we build left invariant star
         products on the $\AN$ group, privileging a pragmatic approach whose geometric
         meaning will be discussed in future work.
         Other invariant star products
         can be obtained from these products using the procedure described in section
         (\ref{IstP}), and even more general star products are given by
         eqs (\ref{genB}, \ref{genPh}).

         The group manifold variables are denoted by $a$ and $n$. The
         infinitesimal left translations are generated by the vector fields
         $\partial_a$ and $\exp[-\,a]\partial_n$. The star product is
         written as:
         \begin{equation} \label{SP}
         (u*v)({\bf x)}= \int {\rm K}[{\bf x},{\bf y},{\bf z}]\,u({\bf
         y})\,v({\bf z})\,d\mu_{\bf y}\,d\mu_{\bf z}\,
         \end{equation} where the (left invariant) measure used is simply
         $d\mu_{\bf x}= da_{\bf x}\ dn_{\bf x}$. To be left invariant, the
         kernel ${\rm K}[{\bf x},{\bf y},{\bf z}]$ has to verify the
         equations:
         \begin{eqnarray} &&\left ( \partial_{a_{\bf x}}+\partial_{a_{\bf
         y}}+\partial_{a_{\bf z}}\right){\rm K}[{\bf x},{\bf y},{\bf z}]=0\qquad ,\\
         &&\left (\exp[-\,a_{\bf x}]\,
         \partial_{n_{\bf x}}+\exp[-\,a_{\bf y}]\,\partial_{n_{\bf y}}+\exp[-\,a_{\bf
         z}]\,\partial_{n_{\bf z}}\right){\rm K}[{\bf x},{\bf y},{\bf
         z}]=0\qquad .
         \end{eqnarray} Hence it depends on four variables instead of six:
         \begin{equation} {\rm K}[{\bf x},{\bf y},{\bf
         z}]=K^L[\a{x}{y},\a{x}{z};\un{y}{x},\un{z}{x}] \qquad ,\label{KL}
         \end{equation} where we have set
         \begin{equation}
         \a{x}{y}:=a_{\bf x}-a_{\bf y}\quad , \quad \un{x}{y}:={n_{\bf
         x}-\exp[-(a_{\bf x}-a_{\bf y})]\,n_{\bf y}} \qquad .
         \end{equation} This condition ensures the left invariance of the star product
         under the $\AN$ group. We now proceed to impose four additional
         conditions: the two that define a star product, i.e. the
         associativity and the existence of right and left unit, as well as
         a condition on the trace and the hermiticity.

         First of all, the existence of a unit element, $u\st LG 1=u$ and
         $1\st LG u=u$, imposes the conditions:
         \begin{eqnarray}
         \int K^L[\a{x}{y},\a{x}{z};\un{y}{x},\un{z}{x}]\, d\mu_{\bf
         z}=\delta ^2[\bf{
         x- y}] &&\label{unitr} \qquad ,\\
         \int K^L[\a{x}{y},\a{x}{z};\un{y}{x},\un{z}{x}]\, d\mu_{\bf
         y}=\delta ^2[\bf{ x- z}] &&\label{unitl} \qquad .
         \end{eqnarray} To fulfill these conditions, we assume that:
         \begin{equation} K^L(\a{x}{y},\a{x}{z};\un{y}{x},\un{z}{x})= \frac
         1{(2\,\pi\,\lambdabar)^2}B(\a{x}{y},\a{x}{z})\,\exp\left \{ i
         \Psi(\a{x}{y},\a{x}{z}; \un{y}{x},
         \un{z}{x})\right\}\label{ampliphas}
         \end{equation} where:
         \begin{equation}\Psi(\a{x}{y},\a{x}{z}; \un{y}{x}, \un{z}{x}):=
         Y(\a{x}{y},\a{x}{z})\,\un{y}{x}+Z(\a{x}{y},\a{x}{z})\,\un{z}{x}
         \quad ,\label{defpsi}
         \end{equation} with $Y(a,\, b)$ and $Z(a,\, b)$ real functions, and $B(a,\, b)$
         complex. This special choice of the phase $\Psi$, linear in the
         $\un yx$ and $\un zx$ variables, as well as the independence of
         the function $B$ in these variables, is dictated by the structure
         of the Fourier transform of the Dirac delta distribution. Eq.
         \ref{unitr} now reads :
         \begin{equation}
         \frac 1{2\,\pi\,\lambdabar\,^2}\int
         B(\a{x}{y},\a{x}{z})\delta[Z(\a{x}{y},\a{x}{z})]\exp\left \{
         iY(\a{x}{y},\a{x}{z})\un{y}{x}\right\} d a_{\bf z} = \delta ^2[\bf
         x-\bf y]\label{unitr2}\qquad .
         \end{equation} To reproduce the distribution of the right hand side of eq.
         (\ref{unitr2}), we have to assume that:
         \begin{equation}
         \delta[Z(\a xy,\a xz)]=\delta[\a xy]/\zeta(\a xz)\quad{\rm i.\
         e.}\quad Z(a,\,b)=0 \quad {\rm iff}\quad a=0\qquad .\label{dZ}
         \end{equation} Hence, to verify eq.(\ref{unitr}), the
         function $B(a,\, b)$ has to satisfy the condition:
         \begin{equation}
         \lambdabar^2 \, \zeta(b)\,\partial_b Y(0,b)= B(0,b)\qquad ,
         \label{unitr3}
         \end{equation} with the sign being fixed by the requirement that $Y(0,b)$ runs
         from $-\infty$ to $+\infty$ when $b$ goes from $-\infty$ to
         $+\infty$. An analogous calculation shows that eq. (\ref{unitl})
         implies:
         \begin{equation}
         \lambdabar^2\,\eta(a)\,\partial_a Z(a,0)= B(a,0)\qquad ,
         \label{unitl3}
         \end{equation} and leads to a relation similar to eq. (\ref{dZ}):
         \begin{equation}
         \delta[Y(\a xy,\a xz)]=\delta[\a xz]/\eta(\a xy)\quad{\rm i.\
         e.}\quad Y(a,\,b)=0 \quad {\rm iff}\quad b=0 \qquad ,\label{dY}
         \end{equation} with $Z(a,0)$ running from $-\infty$ to $+\infty$ when $a$
         varies from $-\infty$ to $+\infty$.

         Let us now impose the following trace condition on the star
         product :
         \begin{equation}
         \int (u \st LG v)({\bf x})d\mu_{\bf x}:=\int k[\b y,\,\b z]u({\bf
         y})\,v({\bf z})d\mu_{\bf y}d\mu_{\bf z}=\int k[\b z,\,\b y]u({\bf
         y})\,v({\bf z})d\mu_{\bf y}d\mu_{\bf z}\qquad ,\label{contr}
         \end{equation} which implies that the two-point kernel:
         \begin{equation}\label{kernel}
         \int K^L[\a{x}{y},\a{x}{z};\un{y}{x},\un{z}{x}]\, d\mu_{\bf
         x}:=k[{\bf y},\,{\bf z}]=k[{\bf z},\,{\bf y}]
         \end{equation}
         is symmetric. We shall not discuss here this condition in all its
         generality, but first restrict ourselves to the special case:
         \begin{equation}\label{kdelta}
         k[\b y,\,\b z]=\delta^2[\b y,\,\b z]\qquad ,
         \end{equation}
         and discuss a slightly more general situation later. After
         integration over $n_{\b x}$ in (\ref{kernel}), this condition may
         be rewritten as:
         \begin{eqnarray} &\frac 1{2\,\pi\,\lambdabar^2}\int
         B(\a{x}{y},\a{x}{z})\delta[Y(\a{x}{y},\a{x}{z})\exp(\,\a xy)+
         Z(\a{x}{y},\a{x}{z})\exp(\,\a xz)]\qquad &\nonumber\\ & \exp\left
         \{ iY(\a{x}{y},\a{x}{z})n_{\bf y}+Z(\a{x}{y},\a{x}{z})n_{\bf
         z})\right\} d a_{\bf x} = \delta(a_{\b y}-a_{\b z}) \,
         \delta(n_{\b y}-n_{\b z})\quad .&\label{tracex}
         \end{eqnarray} To satisfy this relation, we assume that the delta
         distribution appearing in the integrand is equivalent to $\delta
         [\a yx -\a zx]$. In other words, we assume that:
         \begin{equation} Y(a,\,b)\,{\rm e}^{\,a}+Z(a,\, b)\, {\rm e}^{\,b}=0
         \Leftrightarrow a=b\qquad .\label{halfdelta}
         \end{equation} This condition is sufficient to pursue the construction of
         the star product; we shall return to it later.

         Now we analyze the conditions implied by the associativity. The
         star product will be associative if and only if $I_R=I_L$, where:
         \begin{equation} I_R = \int K^L(\b x,\b p,\b y)\,K^L(\b y,\b q,\b r)\, d
         \mu_\b y \qquad {\rm and } \qquad I_L = \int K^L(\b x,\b y',\b
         r)\,K^L(\b y',\b p,\b q)\, d \mu_\b {y'} \qquad .
         \end{equation} After integrations on $n_{\b y}$ and $n_{\b {y'}}$ we obtain:
         \begin{eqnarray} I_R&=&\frac 1{8\,\pi^3\,\lambdabar^4}\int B(\a xp,\a
         xy)\,B(\a yq,\a
         yr){\rm e}^{-i\left\{\left(Y(\a xp,\a xy)\,{\rm e}^{\,\a xp} +Z(\a
         xp,\a xy)\,{\rm e}^{\,\a xy}\right)n_{\b x}\right\}}\nonumber\\
         &&\quad {\rm e}^{i\left\{Y(\a xp,\a xy)\,n_{\b p}+Y(\a yq,\a
         yr)\,n_{\b q} +Z(\a yq,\a yr)\,n_{\b r}\right\}} \nonumber\\
         &&\quad \delta[Z(\a xp,\,\a xy)-Y(\a yq,\,\a yr){\rm e}^{\,\a
         yq}-Z(\a yq,\,\a yr){\rm e}^{\,\a yr}] \ d a_\b y\label{IR}
         \\ &&\nonumber\\ I_L&=&\frac 1{8\,\pi^3\,\lambdabar^4}\int B(\a
         x{y'},\a xr)\,B(\a {y'}p,\a
         {y'}q){\rm e}^{-i\left\{\left(Y(\a x{y'},\a xr)\,{\rm e}^{\,\a
         x{y'}} +Z(\a x{y'},\a xr)\,{\rm e}^{\,\a xr}\right)n_{\b
         x}\right\}}\nonumber\\ &&\quad {\rm e}^{i\left\{Y(\a {y'}p,\a
         {y'}q)\,n_{\b p}+Z(\a {y'}p,\a {y'}q)\,n_{\b q} +Z(\a x{y'},\a
         xr)\,n_{\b r}\right\}} \nonumber\\ &&\quad \delta[Y(\a x{y'},\,\a
         xr)-Y(\a {y'}p,\,\a {y'}q){\rm e}^{\,\a {y'}p}-Z(\a {y'}p,\,\a
         {y'}q){\rm e}^{\,\a {y'}q}] \ d a_\b {y'}\label{IL}
         \end{eqnarray} Using eq. (\ref{halfdelta}), we see that points such
         that $a_\b q
         =a_\b r$ and $a_\b p=a_\b x$ belong to the support of the delta
         distribution appearing in $I_R$, while those such that $a_\b p
         =a_\b q$ and $a_\b r=a_\b x$ belong to the support of the delta
         distribution appearing in $I_L$. To obtain $I_R=I_L$, the supports
         of these delta distributions must coincide (possibly after
         redefinition of the $a_\b {y'}$ variable). To go ahead we assume
         that the support of both delta distributions are located on the
         subset defined by:
         \begin{equation} a_{\b p} - a_{\b q} + a_{\b r} - a_{\b x} =0 \qquad
         .\label{supdel}
         \end{equation} This condition and the requirement of the equality of the
         phases of the two integrals (\ref{IR} and \ref{IL}), leads to the
         set of four equations:
         \begin{eqnarray} Y(\a rq,\a xy)&=&Y(\a {y'}p,\a {y'}q)\label{np}\\ Y(\a yq,\a
         yr)&=&Z(\a {y'}p,\a {y'}q)\label{nq}\\ Z(\a yq,\a yr)&=&Z(\a
         x{y'},\a pq)\label{nr}\\ Y(\a rq,\a xy){\rm e}^{\a rq}+Z(\a rq,\a
         xy){\rm e}^{\a xy}&=& Y(\a x{y'},\a pq){\rm e}^{\a x{y'}}+Z(\a
         x{y'},\a pq){\rm e}^{\a pq}\qquad .\label{nx} \end{eqnarray} To
         ensure the compatibility between the functional relations
         (\ref{np}-\ref{nr}) and eqs (\ref{dZ}, \ref{dY}) we must impose
         the conditions:
         \begin{eqnarray} a_\b y =a_\b x =a_\b p - a_\b q+a_\b r &\Leftrightarrow& a_\b
         {y'}=a_\b q \qquad \label{a1},\\ a_\b y =a_\b r &\Leftrightarrow&
         a_\b {y'}=a_\b p \qquad ,\\ a_\b y =a_\b q &\Leftrightarrow& a_\b
         {y'}=a_\b x =a_\b p+a_\b r - a_\b q \qquad \label{a3} .
         \end{eqnarray} The only linear relation between $a_\b y$ and
         $a_{\b {y'}}$ satisfying relations (\ref{a1})-(\ref{a3})is
         \begin{equation} a_\b {y'}=-a_\b y +a_\b p+ a_\b r\qquad ;\label{ayp}
         \end{equation} the linearity of the relation is imposed by the equality of the
         integrals $I_R$ and $I_L$. Furthermore, by inserting the relations
         (\ref{supdel}, \ref{ayp}) in eqs (\ref{np}, \ref{nr}) we deduce
         that the functions $Y$ and $Z$ depend on one variable only:
         \begin{eqnarray} Y(a,\, b)&=&\tilde Y(b)\quad {\rm and}\quad \tilde
         Y(0)=0\qquad ,\\ Z(a,\, b)&=&\tilde Z(a)\quad {\rm and}\quad
         \tilde Z(0)=0\qquad.
         \end{eqnarray} Moreover eq. (\ref{nq}) implies:
         \begin{equation}
         \tilde Y(a)=\tilde Z(-a)\qquad .
         \end{equation} Condition (\ref{nx}) now becomes:
         \begin{eqnarray}
         \tilde Y(\a pq +\a ry)\ {\rm e}^{\, \a rq}+\tilde Y(\a qr)\ {\rm
         e}^{\,(
         \a pq +\a ry) }&=&\nonumber\\
         \tilde Y(\a pq)\ {\rm e}^{\, \a yq}+\tilde Y(\a qy)\ {\rm e}^{\,
         \a pq}\qquad .\label{LC}&&
         \end{eqnarray} If in this last condition we set $a _\b p= a_\b y$ and $a _\b
         q= a_\b r$, we deduce that $\tilde Y$ and $\tilde Z$ are odd
         functions satisfying:
         \begin{equation}
         \tilde Y(a)=-\tilde Z(a)=\tilde Z(-a)=-\tilde Y(-a)\qquad .
         \end{equation} Finally, considering the condition (\ref{LC}) for $a _\b q=
         a_\b y$ and $a _\b p= a_\b r$, we obtain:
         \begin{equation}
         \tilde Y(2\,a)=2\,\tilde Y(a)\, \cosh (\, a) \qquad ,
         \end{equation} whose only continuous solution, with the condition that $\tilde
         Y(a)$ runs from $-\infty$ to $+\infty$ when $a$ starts from
         $-\infty$, is:
         \begin{equation}
         \tilde Y(a)= \lambdabar^{-1} \sinh( \, a)\qquad ,
         \end{equation} where $\lambdabar$ is a positive constant. These considerations
         allow to fix the $\Psi$-phase of the kernel defining the star
         product as:
         \begin{equation}
         \Psi=\lambdabar^{-1} \left\{ \sinh\left[({a_{\b y}}-{a_{\b
         x}})\right]\, {n_{\b z} }+ \sinh\left[({a_{\b z}}-{a_{\b
         y}})\right]\, {n_{\b x} }+ \sinh\left[({a_{\b x}}-{a_{\b
         z}})\right]\, {n_{\b y} } \right\}\label{phase}\qquad .
         \end{equation} Indeed, it is straightforward to check that this
         expression of $\Psi$
         satisfies the assumption (\ref{supdel}) made about the supports of
         the delta distributions appearing in eqs (\ref{IR} and \ref{IL}).

         Having obtained $\Psi$, we still have to determine the
         $B$-function of the kernel introduced in eq. (\ref{ampliphas}).
         From the existence of right and left units we obtain, using eqs
         (\ref{dZ}, \ref{dY}):
         \begin{equation} \label{BO} B(0,\a xz)=\,\cosh(\,\a xz)\qquad {\rm
         and}\qquad B(\a xy,\, 0)=\,\cosh(\,\a xy)\qquad ,
         \end{equation} while the trace condition (\ref{tracex}) implies the diagonal
         relation:
         \begin{equation} B(\a xy,\,\a xy)=\, \cosh(\,\a xy)\qquad
         .\label{Adiag}
         \end{equation} In addition, the associativity condition requires that the
         $B$-function obeys the quadratic functional equation:
         \begin{equation}
         \frac{B(\a rq,\,\a rq +\a py)\,B(\a yq,\,\a yr)}{\cosh(\a
         qr)}=\frac {B(\a yq,\,\a pq )\,B(\a ry,\,\a rq + \a py)}{\cosh(\a
         qp)}\label{quadasscon}
         \end{equation} from which we easily deduce (by considering the two special
         cases $a_\b q =a_\b r$ and $a_\b p =a_\b y$) that:
         \begin{equation} B(a,b)\,B(b,a)= \cosh(\,a)\cosh(\,b)\cosh(a-b)\qquad .
         \end{equation}

         The last condition that we impose on the star product is the
         hermiticity condition:
         \begin{equation}
         \overline{\left( u\st LG v\right)}=\left( \overline{v}\st LG
         \overline{u}\right)\qquad .\label{hermcon}
         \end{equation} It implies:
         \begin{equation} B(a,\,b)=\overline {B(b,\, a)} \label{amplcon}
         \end{equation} and the antisymmetry of $\Psi$ with respect to this exchange, a
         condition that is already satisfied. So, if we assume the
         hermiticity condition, we obtain:
         \begin{equation} \left | B(\a xy ,\,\a xz) \right |=
         \sqrt{\cosh(\,\a xy)\cosh(\,\a xz)\cosh(\,\a yz)}\qquad
         .\label{ampl}
         \end{equation} The phase $\psi(\a xy,\a xz)$ of the complex function $B$
         (not to be confused with $\Psi$)is an arbitrary odd function of
         two variables, vanishing when one of the variable is an integer
         multiple of $\pi$. Indeed, eqs (\ref{unitr3}, \ref{unitl3}) imply
         that $B(a,0)$ and $B(0,a)$ are real.
         \begin{equation}
         \psi(a,0)=0\qquad.\label{Bphasecon}
         \end{equation} It is a matter of trivial calculations to check that the kernel
         built with the $\Psi$-phase (\ref{phase}) and the function $B$ so
         defined provides a hermitian star product, satisfying the trace
         condition (\ref{contr}) and admitting a left and right unit. Let us
         also notice that both the phase (\ref{phase}) and
         the amplitude (\ref{ampl}) admit a geometrical interpretation in
         terms of geodesic triangles built on the
         three points $\bx$, $\b y$ and $\b z$ (see refs \cite{Pierre2, ZQian}).

         Another left invariant star product under the $\ca \, \cn$ group
         was obtained previously by one of us, starting from completely
         different considerations \cite{Pierre2}. Its phase is also given
         by eq. (\ref{phase}) but $B(\a xy ,\,\a xz)$ is real and given by
         :
         \begin{equation} B(\a xy ,\,\a xz) = \cosh(a_\b y-a_\b z)\qquad .
         \end{equation} This star product does not satisfy the trace condition
         (\ref{contr}) with (\ref{kdelta}) but instead a twisted trace
         condition involving a $K_0$ Bessel function:
         \begin{equation}
         \int (u \st LG v)({\bf x})d\mu_{\bf x}=\frac 1{\pi\,\lambdabar}
         \int K_0\left[ {\lambdabar}^{-1}(n_{\b y}-n_{\b z})\right]
         \delta(a_{\b y}-a_{\b z}) u({\bf y})\,v({\bf z})d\mu_{\bf
         y}d\mu_{\bf z}\quad .
         \end{equation}

         This star product may be reobtained and generalized as follows.
         The phase $\Psi$ given by eq. (\ref{phase}) was built without any
         use of the hermiticity condition. Modulo the computational
         assumptions introduced, its expression results essentially from
         the invariance conditions and one "half'' of the trace condition
         (\ref{halfdelta}) : the condition that leads to the factor
         $\delta(a_\b y -a_\b z)$ on the right hand side of eq.
         (\ref{tracex}). A similar computation as the one described here
         above, but ignoring the other "half" of the trace condition (eq.
         (\ref{Adiag})), leads to:
         \begin{equation} B(a,\,b)\,B(b,\,a)= \frac{B(a,\,a)\,B(b,\,b)}{B(a-b,\,a-b)}
         \cosh^2[(a-b)]\label{Bgen}\qquad .
         \end{equation} Note that by interchanging $a$ and $b$ in this equation we
         obtain that $B(a,\,a)=B(-a,\,-a)$. As a consequence, $B(a,a)$ is
         an even function and the star product satisfies in general a
         twisted trace condition, {\it i.e.} a trace condition that instead
         of $\delta^2[\b y-\b z]$ in eqs (\ref{contr}, \ref{tracex}),
         involves a slightly more general but nevertheless always
         invariant, symmetric kernel $F(n_{\b y}-n_{\b z})\ \delta(a_\b
         y-a_\b z)$ in its right hand side, with:
         \begin{equation} F(n_{\b y}-n_{\b z})=\frac{1}{2\,\pi\,\lambdabar} \int
         B(a,a) {\rm e}^{\frac i{\lambdabar}(n_{\b y}-n_{\b z})\sinh
         a}da\qquad .
         \end{equation} Conversely, if we fix the distribution $F(n_{\b y}-n_{\b z})$,
         $B(a, a)$ must be equal to:
         \begin{equation} B(a, a)=\tilde\cf(a)=\,\cosh(a)\, \hat {\rm
         F}(\lambdabar^{-1}\sinh(a)) \qquad ,\label{Ftilde}
         \end{equation} where $\hat {\rm F}(k)=\int F(n)\exp(-i\,k\,n)\, dn $ is the
         Fourier transform of $F$.

         The most general solution of eq. (\ref{Bgen}) can easily be
         expressed by decomposing the function $B(a,b)$ into its symmetric
         $B_s$ and antisymmetric $B_a$ parts:
         \begin{equation} B_s(a,b)=\frac12\left(B(a,b)+B(b,a)\right)\qquad ,\qquad
         B_a(a,b)=\frac12 \left(B(a,b)-B(b,a)\right)\qquad .
         \end{equation} Equations (\ref{B0})imply that
         the function $B_a$ vanishes when one of its arguments is zero
         ($B_a(a,0 )=0$), but otherwise is arbitrary. The symmetric part of
         the $B$-function depends on the function $\tilde\cf(a)$ defined in
         eq. (\ref{Ftilde}), with the additional condition $\tilde\cf
         (0)=1$ ensuring that $\int_{-\infty}^\infty F(n) dn=1$. It is
         given by:
         \begin{equation}
         B_s^2(a,b)=\frac{{\tilde\cf}(a){\tilde\cf}(b)} {{\tilde\cf}
         (a-b)}\cosh^2(a-b)+B_a^2(a,b)\label{amplsymgen}
         \end{equation} Choosing $B$ symmetric and real with
         $\tilde\cf(\alpha)= 1$ corresponds to the star product presented
         in \cite{Pierre2}, $\tilde\cf(\alpha)=\,\cosh \alpha $ to a
         hermitian star product such that the trace condition is given by
         eqs. (\ref{contr}) and (\ref{kdelta}).

         Let us also mention that some of the star products we have
         discussed here above are easily related to the Moyal-Weyl star
         product (denoted $*$)\footnote{Formally, all star products are
         equivalent to a Moyal-Weyl star product; the transformation $T$
         that makes the correspondence can always be constructed, step by
         step, as a formal series. The point here is that we have explicit
         transformations that allow to define the functional space on which
         the star product constitutes an internal composition law.}:
         \begin{equation} (u* v)(\b x):=\frac 1{(2\pi\,\lambdabar)^2} \int {\rm
         e}^{\frac{i}{\lambdabar} (\b x-\b y)\wedge (\b x -\b z)}u(\b y)\,
         v(\b z)\, d\mu_\b y d\mu_\b z\label{stM} \qquad ,
         \end{equation} via the sequence of transformations:
         \begin{equation} (u\st LG v)=T^{-1}\left[T[u]* T[v]\right]\label{stTM}
         \qquad ,
         \end{equation} where:
         \begin{equation} T[u](a,\, n):=\frac {1}{2\,\pi\,\lambdabar}\int {\rm
         e}^{-\frac i{\lambdabar}\xi\,n}\cp(\xi){\rm e}^{\frac
         i{\lambdabar} \sinh(\xi)\nu}u(a,\, \nu)\, d\nu\, d\xi \qquad ,
         \label{Ttransf}
         \end{equation} slightly generalizing, by an extra multiplication by the
         (non-vanishing) complex function $\cp$, a similar transformation
         first obtained in ref. \cite{Pierre2}. The kernel of the star
         product so defined is given by:
         \begin{equation} \label{noyauP}
         K^L[\a{x}{y},\a{x}{z};\un{y}{x},\un{z}{x}]=\frac{1}{(2\,\pi\,\lambdabar)^2 }
         \frac{\cp(\a
         yx)\cp(\a xz)}{\cp(\a yz)}\cosh(\a yz)\, \exp(i\Psi)\qquad .
         \end{equation}

         Formulas (\ref{stTM}, \ref{Ttransf}) clarify some of the
         constraints imposed to the function $\cp$: $\cp(0)=1$ is necessary
         to obtain $u\st LG 1=1\st LG u=u$, as otherwise we would have
         $u\st LG 1=1\st LG u= \cp(0)\ u$; its positivity on the real axis
         is necessary for the existence of $T^{-1}$, and it is only if
         $\cp(a)=\overline{\cp(-a)}$ that the hermiticity condition
         (\ref{hermcon}) can be fulfilled.

         A tedious but elementary calculation shows that all the star
         products considered here may be seen as deformations of the
         canonical symplectic structure defined by the surface element
         under the $\ca \cn$ group (if $\cp(0)$ is properly normalized).
         Indeed, using the kernel (\ref{ampliphas}), one finds at first
         order in $\lambdabar$ :
         \begin{eqnarray} \label{firstord} (u\st LG v)(\b x)&=&u(\b x)\,v(\b
         x)-i \,\lambdabar
         \, B(0,0)\, u(\b x)(\,\lpartial_{a_{\b x}}\rpartial_{n_{\b
         x}}-\lpartial_{n_{\b x}}\rpartial_{a_{\b x}})v(\b x) \nonumber\\
         && + i \,\lambdabar
         \,(v(\b x) \partial_{n_{\b x}}u(\b x) B^{(0,1)}(0,0) -
         u(\b x) \partial_{n_{\b x}}v(\b x) B^{(1,0)}(0,0)) + O(\lambdabar ^2) \, .
         \end{eqnarray}
         In order to define an invariant star product, the amplitude
         $B(a,b)$ has in particular to satisfy the relations (\ref{BO}),
         which force $B(0,0) = 1$ and $(\partial_a B)(0,0)=(\partial_b
         B)(0,0)=0$. Equivalently, this implies that $\cp(0)=1$ in the
         kernel (\ref{noyauP}).

         To complete the construction of the star products,
         the functional space on which they
         are defined must be specified. Let us focus on the star
         products defined on $\ca \cn$ which are related to the Moyal-Weyl
         product by an explicit intertwiner $T$, given by eqs. (\ref{stTM},
         \ref{Ttransf}). We remind that the Schwartz space $\cs$ is stable
         with respect to the Moyal-Weyl star product \cite{Moy}. The
         star product given by (\ref{stTM}) will consequently define an
         associative algebra structure on the space $T^{-1}\cs \subset
         \cs'$, provided $T[f]
         \in \cs\ \forall \ f\in \cs$. As the Fourier transform is an isomorphism
         of the space $\cs$, this will be the case if the
         function $\cp(a)$ is $C^\infty$, nowhere zero on the real axis
         and increases at infinity not faster than a power of $\exp(|a|)$. Unfortunately
         the functional space so obtained does not yet contain the constant
         function $1$.

         To overcome this difficulty, we forget, for a while, about the
         dependence of eq. (\ref{Ttransf}) on the
         variable $a$ and limit ourselves to the $n$ dependence of
         the function. The mapping $T^{-1}$:
         \begin{equation}
         T^{-1}[a,\,u](\nu):=\frac {1}{2\,\pi\,\lambdabar}\int
         {\rm e}^{-\frac
         i{\lambdabar} \sinh(\xi)\,\nu}\cp^{-1}(\xi)
         {\rm
         e}^{\frac i{\lambdabar}\xi\,n}\cosh(\xi)\ u(a,\,n)\, dn\,
         d\xi \qquad , \label{InvTtransf}
         \end{equation}
         is well-defined as a linear
         injection of $\cs_{(n)}$ (the Schwartz space of functions of the $n$
         variable) into the
         tempered distribution space
         $\cs'$ \cite{Pierre2}, and also as an operator $T^{-1}: \cs' \mapsto
         \cs'$. Consider now the space $\cb_{(n)}$ of smooth
         bounded functions with all their derivatives bounded in the
         variable $n \in \mathbb{R}$. This space can be seen as a subspace of $\cs'$.
         Defining $\ce_{(n)}=T^{-1}[\cb_{(n)}]$, we obtain a deformed algebra
         containing the constants. Moreover, because $T$ only affects the $n$-variable,
         $\ce_{(n)}$ also contains the bounded functions in the $a$-variable.
         Another way of implementing the constants in our deformed algebra is to
         consider the unitalization $\CC \oplus T^{-1} \cs $.

         Finally, let us notice that if we replace in the transformation
         (\ref{Ttransf}) used in eq.(\ref{stTM}) the $\sinh$ function by an
         arbitrary monotone, real, odd function $\Phi$, running from
         $-\infty$ to $+\infty$, and in the $B$-function the $\cosh$
         functions by the derivative $\Phi'$ of this function, we still
         obtain the kernels of hermitian star products, which are however
         in general no longer left invariant under $\ca \cn$. In
         particular the quadratic associativity condition
         (\ref{quadasscon}) remains satisfied as well as the trace
         condition if the $B$-function is built according to eq.
         (\ref{ampl}). Even more general expressions of the kernel $K[\b x, \b
         y,\b z]= \frac {1}{4 \pi^2}B\,\exp[i\Psi]$ that provide associative
         star products admitting 1 as left and right unit and verifying a trace
         condition do exist. For instance, we have:
         \begin{eqnarray}
         B[\b x, \b y,\b z]
         &=&\sqrt{\Phi'(0)\Phi'(a_{\b x}-a_{\b y})\Phi'(a_{\b y}-a_{\b
         z})\Phi'(a_{\b z}-a_{\b x})}\phi(a_{\b y})\,\phi(a_{\b z})\qquad
         ,\label{genB}\\
         \Psi[\b x, \b y,\b z]
         &=&n_{\b
         x}\Phi(a_{\b y}-a_{\b z})\,\phi(a_{\b x})+n_{\b y}\Phi(a_{\b z}-a_{\b
         x})\,\phi(a_{\b y})+ n_{\b z}\Phi(a_{\b x}-a_{\b y})\,\phi(a_{\b
         z})\qquad
         .\label{genPh}
         \end{eqnarray}
         In particular, choosing the functions $\Phi(a)=\sinh(a)$ and
         $\phi(a)=\exp(a)$ leads to the right invariant star product.

         \subsection{Star products in $AdS_3$ and $\EBTZ$}

         In the previous section we have obtained star products on the
         $\AN$ group manifold. From section (\ref{IstP}) we know how to
         construct induced star products on spaces on which this group
         acts, in particular on the fibers $\b S_0$ of $AdS_3$ and
         $\Sigma_0$ in $\EBTZ$, as far as the appropriate functional spaces
         are specified. Of course the $\AN$ group also acts on $AdS_3$ and
         $\EBTZ$ [see section \ref{Geom}], and as consequence induces star
         products on them. The value of the star product of two functions
         defined on these spaces, at the point of coordinates
         $[\rho_0,\phi_0,w_0]$, is simply given by the star products of the
         same functions considered as functions on the D-brane
         $\rho=\rho_0$, at
         the point of coordinates $[\phi_0,w_0]$

         We now give detailed formulas for the star products induced on $\b U$ and
         $\EBTZ$. For similar formulas on the D-brane $\b S_0$ of $AdS_3$
         and even on the D-brane $\Sigma_0$ in $\EBTZ$
         [see section (\ref{Branes})] it is sufficient to forget about the $\rho$
         dependence here below.

         The composition law on the $\AN$ two parameter group, is
         given by:
         \begin{equation} [a_1,\,n_1]*[a_2,\,n_2]=[a_1+a_2,n_2+n_1\,\exp(-a_2)]
         \end{equation} from which we infer:
         \begin{equation} [a,\,n]^{-1}=[-a,\,-n\exp a\,]\qquad .
         \end{equation}
         In the following, when no confusion could arise, we shall identify
         points of $\b U$, $\EBTZ$ or group elements of $\AN$,
         respectively, with their triplet of coordinates or pair of
         parameters.

         The left action of an element $(a,\,n)$ on a point is
         given by the mapping:
         \begin{equation} [\rho,\, \phi,\,w]\mapsto [\rho,\,\phi + a,\,w
         +n\,\exp(-\phi)]\qquad ;\label{Lac}
         \end{equation} the right action is given by:
         \begin{equation} [\rho,\, \phi,\,w]\mapsto [\rho,\,\phi + a,\,w\,\exp(-a)
         +n]\qquad .\label{Rac}
         \end{equation}

         Let us suppose to have a left invariant star product (\ref{SP}) on the group
         $\AN$, defined by a kernel $K^L$ (eq. \ref{KL}). It induces [see eq.
         (\ref{babel})] on the domain $\b U$
         the star product:
         \begin{eqnarray} (U\st{r}{\b U} V)[\rho,\,\phi,\,w]&
         =&\int
         K^L[-a_1,-a_2,\,n_1,\,n_2]\,U[\rho,\,\phi-a_1,\,w-n_1\,\exp(-\phi
         +a_1)]\nonumber\\ && V[\rho,\,\phi-a_2,\,w-n_2\,\exp(-\phi
         +a_2)]\,da_1\,dn_1\,da_2\,dn_2\qquad ,\\
         &=&\int
         K^L[-\phi+\alpha_1,-\phi+\alpha_2,\,(w-\nu_1)\exp(\alpha_1),\,(w-\nu_2)\exp(\alpha_2)]\,U[\rho,\,\alpha_1,\,\nu_1]\nonumber\\
         &&
         V[\rho,\,\alpha_2,\,\nu_2]\,\exp(\alpha_1)\,d\alpha_1\,d\nu_1\,\exp(\alpha_2)\,d\alpha_2\,d\nu_2\qquad
         ,
         \end{eqnarray}
         which is invariant under the right action (\ref{Rac}) on $\AN$ on $\b U$.
         On the other hand we also obtain, from the kernel $K^L$, a
         right invariant star product on the group $\AN$:
         \begin{eqnarray} (u\st{R}{\AN}v)[a_x,\,n_x]&=&\int
         K^L[-a_x+a_y,-a_x+a_z,\,(n_x-n_y)\exp(a_y),\,(n_x-n_z)\exp(a_z)]\nonumber\\
         &&u[a_y,\,n_y]
         \,v[a_z,\,n_z]\,\exp[a_y]\,da_y\,dn_y\,\exp[a_z]\,da_z\,dn_z
         \end{eqnarray} and, as a consequence, another induced star product on $\b U$:
         \begin{eqnarray} (U\st{l}{\b U}V)[\rho,\,\phi,\,w]& =&\int
         K^L[a_1,a_2,-\,n_1\,\exp(a_1),-\,n_2\,\exp(a_2)]\,U[\rho,\,\phi-a_1,\,(\,w-n_1)\exp(a_1)]\nonumber\\
         &&
         V[\rho,\,\phi-a_2,\,(w-n_2)\exp(a_2)]\,\exp(a_1)\,da_1\,dn_1\,\exp(a_2)\,da_2\,dn_2\\
         & =&\int
         K^L[\phi-\alpha_1,\phi-\alpha_2,\,\nu_1-w\exp(\phi-\alpha_1),\,\nu_2-w\exp(\phi-\alpha_2)]\,U[\rho,\,\alpha_1,\,\nu_1]\nonumber\\
         &&
         V[\rho,\,\alpha_2,\,\nu_2]\,d\alpha_1\,d\nu_1\,d\alpha_2\,d\nu_2\qquad .
         \end{eqnarray}
         This expression is invariant under the left action of $\AN$ on $\b U$ and
         is obviously compatible with the quotient leading to the
         $\EBTZ$ space. Indeed using the global coordinate system
         (\ref{metmax}) the identification yielding the maximally extended
         space
         from $\b U$ reads as:
         \begin{equation}
         [\rho,\, \phi,\,w]\equiv [\rho,\,\phi + 2\pi\,\sqrt{M},\,w]\qquad .
         \end{equation}
         Therefore, if the functions $U$ and $V$ are periodic on $\b U$,
         in the $\phi$ variable, the star product $U\st{l}{\b U}V$ will
         also be periodic, contrary to what happens with the right invariant
         star product.

         This allows us to define a deformed product {\itshape at the quotient
         level}, i.e. the star-product of two functions $U$ and $V$ on
         $\EBTZ$. To this end we adopt a method of images, that is, we extend
         periodically the function on $\b U$ by defining
         \begin{equation}
         \tilde{U}[\rho,\phi,w] = \sum_{k \in \ZZ} U [\rho,\phi + k 2\pi\,
         \sqrt{M},w]
         \end{equation}
         and
         \begin{equation}
         U \underset{\widetilde{BTZ}}{\star} V \quad = \quad \tilde{U} \st{l}{\b U}
         \tilde{V} \quad .
         \end{equation}

         To give a precise meaning to this formal sum, one proceeds as
         follows. Denoting by $\pi : \b S_0 \mapsto \Sigma_0 $ the quotient
         map onto the $\EBTZ$ space. If $C^\infty_c$ is the space of smooth
         compact supported functions on $\Sigma_0$, we notice that $\pi
         ^\star (C^\infty_c (\Sigma_0)) \subset \ce $
         is the space of smooth $\phi$ periodic functions which, for fixed
         value of $\phi$, have compact support in $w$. Now we observe that
         the mappings $T$ and $T^{-1}$ only involve the $w$ variable,
         $\phi$ being a spectator variable. Knowing that the space
         $\cb_{(\phi,\,w)}$ of two-variable ${(\phi,\,w)}$ functions is
         stable under the Moyal-Weyl product (\ref{stM})\cite{Rief}, we
         find that if $U,\, V\in
         \ce_{(\phi,\,w)}=T^{-1}[\cb_{(\phi,\,w)}]$, $U\st l{\b U} V$ is
         also in $\ce_{(\phi,\,w)}$. The subalgebra of $\ce_{(\phi,\,w)}$
         generated by $\pi^\star (C^\infty_c(\Sigma_0))$ is then
         constituted by $\phi$-periodic functions, owing to the
         $\ca$-invariance of the star product. It is therefore identified
         with a function algebra on $\Sigma_0$ and on the $\EBTZ$ space.

         \section{Deformation of Dirac operators}\label{Dirac}

         Motivated by Connes' definition of noncommutative spectral triples
         \cite{ConLan}, we are interested in the definition of Dirac
         operators associated to the deformed algebras introduced in
         sections 4.2 and 4.3 .
         Note that the left and right action of the
         group $\AN$ are crucially different. The left actions constitute
         an isometry group for the metric (\ref{metmax}) or its restriction
         to the D-brane $\rho =\rho_0$. The Dirac operator on this D-brane
         can be written as
         \begin{equation}
         \dirac_{\rho_0}=\mbox{sech}\rho_0 \left[\left(
         \begin{array}{cc}
         0&\partial_\phi-w\,\partial_w+\partial_w\\
         \partial_\phi-w\,\partial_w-\partial_w&0
         \end{array}
         \right) -\frac{1}2 \left(
         \begin{array}{cc}
         0&1\\
         1&0
         \end{array}
         \right)\right]\qquad .\label{Dr}
         \end{equation} Two vector fields appear naturally in this operator:
         $\delta_\pm=\partial_\phi-w\,\partial_w\pm\partial_w$. They
         constitute left invariant (isometry invariant) vector field and
         are the generators of the right transformations. If $\bf{\Psi}$ is
         a two component spinor field on the D-brane and $u$ a function,
         the Dirac operator (\ref{Dr}) behaves as a derivative in the
         star product algebra:
         \begin{equation} \dirac_{\rho_0}\left(u\st{r}{\b S_0}\b
         \Psi\right)=\left(\dirac_{\rho_0}u\Id\right)\st{r}{\b S_0}\b
         \Psi+u\st{r}{\b S_0}\dirac_{\rho_0}\b \Psi\qquad ,\label{Leib1}
         \end{equation}
         the components of the star product of a spinor field and a
         function being defined as the star product of the function with
         each of the components of the spinor field. Accordingly the
         commutator $[\dirac_{\rho_0},\,u]=\dirac_{\rho_0}u\Id$ is a
         bounded operator for all Schwartz functions on $\b S_0$. This
         way, we have at hand the ingredients for a noncommutative spectral
         triple on $AdS_2$ and $AdS_3$ spaces. Nevertheless, as discussed
         in the preceding section, this right invariant star product is
         {\itshape not} compatible with the quotient leading to the BTZ
         spaces.

         Turning to $\EBTZ$ spaces,
         an immediate calculation shows that the operator (\ref{Dr}) does
         not define a derivation on the algebra based on $\st{l}{\b S_0}$.
         Nevertheless we can provide a deformation of the Dirac operator
         in an algebraic framework, analogous to those considered in ref. \cite{Wata}
         We shall consider genuine deformations
         of $\dirac$, built as follows. Derivative in the directions given
         by left invariant vector fields obey the Leibnitz rule with
         respect to the star product $\st{r}{\b S_0}$ only because this
         star product was itself right invariant. Accordingly, right
         invariant vector fields will define derivatives in the algebra
         obtained from the left invariant star product $\st{l}{\b S_0}$.
         The point here is that these derivative operators read as:
         $\mu_1=\partial_\phi$ and $\mu_2=\exp(-\phi)\partial_w$ and the
         Dirac operator expresses itself as:
         \begin{eqnarray}
         \dirac_{\rho_0}&=&\mbox{sech} \rho_0\left[ \left(
         \begin{array}{cc}

         0&1\\

         1&0

         \end{array}
         \right)\partial_\phi+\left(
         \begin{array}{cc}

         0&(1-w)\exp(\phi)\\

         -(1+w)\exp(\phi)&0

         \end{array}
         \right) \exp(-\phi)\partial_w
         \right. \nonumber\\
         &&-\frac{1}{2} \left. \left(
         \begin{array}{cc}

         0&1\\

         1&0

         \end{array}
         \right)\right]\\
         &:=&\Gamma^I[\phi,w]\,\mu_I+K[\phi,w] \qquad .\label{DrR}
         \end{eqnarray}

         Actually the matrix $K$ is constant (with respect to the $\phi$
         and $w$ variables), as well as $\Gamma^1$, but we may consider
         slightly more general situation than the simplest Dirac operator
         considered here, for instance by adding gauge field coupling. The
         idea to extend the Dirac operator as an operator
         (denoted $\Ddirac$), which on the module of functions and spinor fields
         with the star product as multiplication enjoys similar properties as
         a derivative operator,
         consists in defining it as
         follows:
         \begin{equation}
         \left(\Ddirac\Psi\right)^j=\mu_I(\Psi^k)\st{l}{\b
         S_0}\left(\Gamma^I\right)^j_{\
         k}+\Psi^k\st{l}{\b S_0}\left(K\right)^j_{\ k}\qquad .
         \end{equation}
         We have ordered the various terms and multiply them using the
         star product, so that a formula analogs to eq. (\ref{Leib1})
         remains valid :
         \begin{equation} \left( \Ddirac_{\rho_0}(u\st{l}{\b S_0}
         \Psi)\right)^l= u \st{l}{\b S_0} (\Ddirac_{\rho_0} \Psi)^l +
         \mu_I(u) \st{l}{\b S_0} \Psi^k \st{l}{\b S_0} (\Gamma^I)^l_k
         \end{equation}
         The motivation to push the non constant $\Gamma$ matrices to the
         right is the requirement that the Clifford multiplication remains
         (left) linear. With this definition
         $\left([\Ddirac_{\rho_0},\,u]\Psi\right)^l=\mu_I(u)\st{l}{\b
         S_0}\Psi^k \st{l}{\b S_0} (\Gamma^I)^l_k$. This operator is not
         bounded and consequently another deformation of the Dirac operator
         should be used in Connes' construction. Note that, while the
         vector field $\mu_2$ is not defined on $\EBTZ$ due to the
         appearance of the factor $\exp(\-\phi)$, the Dirac operator (of
         course) {\bf and} its twisted version $\Ddirac_{\rho_0}$ are well
         defined.

         \section{Discussion}
         BTZ black holes exhibit interesting geometrical structures. As
         shown in \cite{BRS}, $AdS_3$ space admits a foliation by
         twisted conjugacy classes in $\SL$, stable under the
         identification subgroup leading to the the massive non-rotating
         black hole. Each of the leaves of the foliation showed to be
         canonically endowed with a Poisson structure. This essentially
         reduces the study of the black hole to a two-dimensional problem.
         According to \cite{BachPetr}, these two-dimensional conjugacy
         classes represent D-strings in $AdS_3$ space-time. Consequently,
         they project into closed branes in the BTZ black hole background,
         whereas a generic D-string in $AdS_3$ projects onto an infinite
         D-string wound around the black hole. We analyzed these closed
         D-strings in section \ref{Branes}. using the (approximate)
         space-time description given by the DBI action.

         Furthermore, the Poisson structure intrinsically defined on each
         leaf is the primary input for the construction of a noncommutative
         deformation of the usual pointwise product of functions (i.e. a
         star product). In view of this construction, we first showed that the
         non-rotating massive BTZ space-time admits two nonequivalent
         extensions beyond the singularities in the causal structure, much
         in the same way as the two-dimensional Misner space can be maximally
         extended. In one of these extensions, each of the leaves of the
         foliation can be seen as orbits of the action of the non-Abelian
         two parameter subgroup $\AN$ of $\SL$. This action allowed us to
         construct a family of induced star products on the two-dimensional
         leaves of $\widetilde{BTZ}$ and on $AdS_3$. This result extends
         Rieffel's strict deformation theory for manifolds admitting an
         action of Abelian groups to manifolds with a non-Abelian group
         action, and was first introduced in \cite{Pierre2} using
         different techniques.

         This construction raises some questions about
         causality. Indeed, the closure of the support of the non-formal
         star product of two compactly supported functions generally
         extends beyond the closure of the intersection of the supports of
         the functions. However, for the star products
         satisfying the trace condition (\ref{kdelta}),
         we get the weaker condition that the integral of the
         star product of two functions always vanishes if their supports do not
         overlap; otherwise, the star product depends only on the values of the
         functions on the intersection of their supports. Of course, we
         recover causality if we restrict ourselves to the {\itshape
         formal} star products, obtained from (\ref{stTM}) by extending the
         expansion (\ref{firstord}) to all orders.

         There are, in our opinion, two main directions worth being
         investigated. The first is related to Connes' construction (of
         Yang-Mills type actions). We showed in section 5 that we have at hand the first ingredients
         needed for a non-commutative spectral triple on $AdS_2$
         and $AdS_3$ (see \cite{Stro,Mor} for the definition of
         semi-Riemannian spectral triples).

         The second direction results from the close link between string theory (branes
         and open strings) and non-commutative geometry. In this paper, we
         focused on the
         {\itshape approximate} space-time approach of the D-branes wrapping
         around the BTZ black hole,
         using the DBI action. Actually, in the {\itshape exact}
         world-sheet approach, a D-brane is described in boundary conformal
         field theory by its interaction with closed strings, i.e. by a
         closed string state called boundary state. For branes in a flat
         background with constant B-field, it was shown in \cite{SchomD}
         (see also \cite{Doug, Chu}) from the operator product
         expansions of tachyonic open string vertex operators that the
         brane's world-volume geometry is given by a Moyal-Weyl deformation
         of the classical algebra of functions on the brane, and scattering
         amplitudes of massless open string modes give rise to a
         non-commutative Yang-Mills theory \cite{SeibWitt,BranesCurved}.
         Branes in compact group manifolds reveal a more intricate
         structure, but non-commutative geometry also emerges due to the
         presence of the WZW background B-field (see for instance
         \cite{BraneDyn}). It could be interesting to explore the possible
         link between the family of star products we introduced and the
         worldvolume geometry of D-branes in the non-compact $\SL$ WZW
         model.

         \acknowledgments
         S.D., Ph.S. and M.R. acknowledge support from the
         Fonds National de la Recherche scientifique
         through an F.R.F.C. grant.

         \end{document}